\documentclass{mem}
\usepackage{natbib}
\usepackage{txfonts}
\usepackage{balance}
\usepackage{graphicx}
\usepackage{flushend}
\usepackage[a4paper,breaklinks,pdftex]{hyperref}
\idline{75}{282}
\begin{document}

\title{
Building the high-redshift Hubble Diagram with quasars
}

   \subtitle{}

\author{
Matilde \,Signorini\inst{1,2} 
\and Guido \, Risaliti\inst{1,2}
\and Andrea \, Sacchi\inst{3}\
\and Elisabeta \, Lusso\inst{1,2}\  
\and Emanuele \, Nardini\inst{2}
  }

\institute{
Università degli Studi di Firenze -- Via Sansone 1, Sesto Fiorentino, 50019, Firenze, FI, Italy
\and
INAF - Osservatorio Astrofisico di Arcetri --  Largo Enrico Fermi, 5, 50125 Firenze, FI, Italy
\and
IUSS Pavia -- Palazzo del Broletto, Piazza della Vittoria, 15, 27100 Pavia PV, Italy\\
\email{matilde.signorini@unifi.it}
}

\authorrunning{Signorini }

\titlerunning{Building the high-redshift Hubble Diagram with quasars}


\abstract{ In recent years, quasars have been shown to be reliable standardizable candles, thanks to the non-linear relation between their X-rays and ultraviolet luminosity. Quasars are also very numerous and they are found at all the observed redshifts: this allows us to investigate the expansion rate of the Universe and the cosmological parameters in a previously almost untested redshift range ($z\sim2-7$). At redshift higher than 1.5, the Hubble Diagram of quasars shows a significant tension with the predictions of the $\Lambda$CDM model.\\ I will show how detailed optical/UV and X-rays spectroscopic analysis can be used (i) to obtain more precise distance estimates, and (ii) to derive information about the physical process behind the luminosities relation, and discuss the cosmological implementations.
}
\maketitle{}

\section{Introduction}
One of the  most important tools for cosmology is the distance-redshift relation, the ``Hubble Diagram". Its shape depends on the composition of the Universe and on its expansion history. Therefore, we can use it to test cosmological models or, inside a given model, to put constraints on the cosmological parameters. To build the Hubble diagrams, we need a sample of objects for which both the redshift and the distance are known, in a cosmology-independent way. To satisfy the latter requirement, we can use objects that are standard candles, that is, objects for which we know the intrinsic luminosity and we can therefore derive the luminosity distance from the flux-luminosity relation $F=\frac{L}{4\pi D_L^2}$. The most famous standard candles are Supernovae Ia, for which the intrinsic luminosity can be derived from the luminosity evolution after maximum light thanks to the Phillips relation \citep{Phillips93}. With them, it is possible to build the Hubble Diagram from redshift $z=0$ to $z\sim1.5$, up to when the Universe was $\sim$6 billion years old. The SNIa Hubble Diagram shows that the expansion of the Universe is
accelerating \citep{Riess98, Perlmutter99}. \\
However, the Hubble Diagram of Supernovae Ia leaves the first part of the Universe expansion history not-investigated. This is where quasars can come at hand: they are numerous, very luminous and observed up to redshift $\sim$7, when the Universe was less than a billion years old \citep{Banados16}. We usually define ``quasar" an Active Galactic Nucleus (AGN) with a bolometric luminosity that significantly exceeds the one of the whole host galaxy. Quasars are not standard candles, as their bolometric emission can vary up to orders of magnitude ($10^{11}-10^{14} L_{\odot}$) between one object and another. However, there is one observational evidence that can let us overcome this: the presence of a non-linear relation between the X-ray and the UV luminosities in quasars. This is usually parametrised as a linear relation in the logarithmic space:
\begin{equation}
    \log(L_X) = \gamma \log(L_{UV}) + \beta
\end{equation}
where $\gamma$ and $\beta$ are the slope and the normalization of the relation, respectively. If we substitute in this equation the relation between luminosity and flux, we can derive the luminosity distance $D_L$:
\begin{equation}
    \log(D_L) = \frac{1}{2-2\gamma} (\log(f_X) - \gamma\log(f_{UV})) + \beta'
\label{rel}
\end{equation}

This relation was known since the '80s, but it had not been implemented in cosmology due to the high observed dispersion that affected it. 

\begin{figure}
\centering
\includegraphics[scale=0.38]{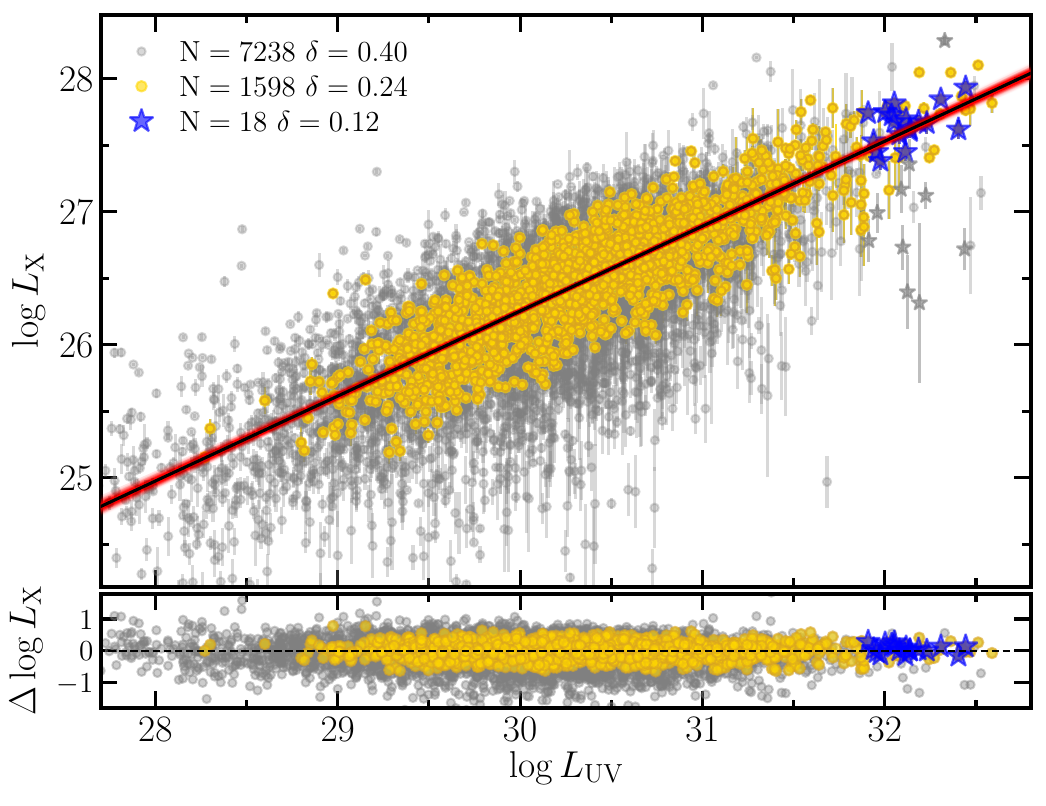}
\caption{Non-linear relation between X-ray and UV luminosities in quasars. Different colors mark different quality subsamples. Figure from \cite{RL19}.}
\label{fig:lxluv}
\end{figure}

In Figure \ref{fig:lxluv} we can see it in the logarithmic space; the grey points represent a sample of $\sim7000$ quasars, for which the dispersion around the best fit of the relation is $\delta = 0.40$ dex. Such a high dispersion makes the distance estimates basically useless, because they would be affected by incredibly high uncertainties.\\
In the past years, it has been shown that most part of this dispersion is not intrinsic, but it is due to observational causes that can for the most part be removed if we only select objects with ``good quality" data, that are not affected by dust reddening, gas absorption or other kinds of biases. More details can be found in \cite{Lusso16}, \cite{RL19}. In Figure \ref{fig:lxluv}, the yellow points represent this ``cleaned" quasar sample, and the dispersion is reduced to $\delta = 0.24$ dex. With this dispersion, it is possible to actually build a Hubble Diagram for quasars. In order for such a Hubble Diagram to be reliable, however, we need the relation parameters $\gamma$ and $\beta$ not to be varying with the redshift. Regarding the slope parameter $\gamma$, we can test this by dividing the quasar sample into small redshift bins and, inside each bin, testing the relation using fluxes instead of luminosities. This has been done in many previous works \citep[e.g.,][]{Lusso17, RL19, Bisogni21}, and the slope $\gamma$ has always been found to be constant with the redshift. Regarding the parameter $\beta$, being it a normalization parameter, we have no way to directly test it being constant with the redshift in a cosmology-independent way. However, we can use the common redshift range of SNIa and quasars to cross-correlate the quasars Hubble Diagram and determine the parameter $\beta$. As discussed in previous works, the fact that in the common redshift range quasars and SNIa perfectly overlap once the cross-normalization parameter is set means that $\beta$ is not changing with the redshift either \citep[for more details, see ][]{RL19}.\\ It is important to notice that, because we are using SNIa to calibrate the quasars Hubble Diagram, we can not derive an independent measure of the Hubble constant $H_0$ from quasars.\\

Including quasars, the Hubble Diagram is extended up to much higher redshifts, when the Universe was around one billion years old. 
With this extension, a $\sim$4$\sigma$ tension is found between the data (SNIa + quasars) and the predictions of the $\Lambda$CDM model as derived from the Planck collaboration \citep{Planck16}, widely accepted as the standard model for cosmology. This tension starts to build up at a redshift higher than 1.5, so it can not be seen using SNIa alone. At the same time, there is a perfect match between quasars and SNIa in the common redshift range, which gives reliability to results obtained with quasars at higher redshifts \citep{RL19}. This tension is usually  tested by fitting the quasar+SNIa Hubble Diagram with a cosmographic function, and then comparing the best-fit parameters of the function with what they would be supposed to be in a given cosmological model. More details about this can be found, for example, in \cite{Bargiacchi21}, where a $>4\sigma$ tension is determined. Possible bias and systematics that may be causing this tension are thoroughly discussed in \cite{Lusso20} (Section 9), and they alone can not explain such a significant tension. \\Tests of alternative cosmological models are also being carried out; in the realm of models in which the Dark Energy (DE) component evolves with time, quasars suggest an increasing ("phantom") DE \citep[for details, see for example ][]{Bargiacchi22}.\\
One of the limits of quasars implementation as cosmological probes is the still-quite-high observed dispersion of the luminosities relation. Another still-persisting issue is the fact that there is no clear physical model that explains the luminosities relation. We know that this relation must come from the interaction between the accretion disc and the X-ray Corona. We also know that an energy transfer between the disc and the Corona must exist, otherwise, there could not be a persistent X-ray emission, as the hot electrons that upscatter UV photons would rapidly lose their energy. At the same time, the exact nature of the interaction is still an open question.\\

\section{UV spectral analysis}
I analysed the UV spectra of a sample of 1761 quasars, coming from the \cite{Lusso20} sample. Given that we are interested in the cosmological implementation of quasars, we selected only objects with a redshift higher than 0.5. As discussed in \cite{Lusso20} (Section 5.1 and 8), 2500{\AA} monochromatic luminosities may not be so reliable at lower redshift because of contamination from the light of the host galaxy. Having $z > 0.5$ as the lower limit still gives us a significant redshift range ($\sim$ 0.5-1.2) for the cross-calibration of quasars with SNIa. \\With this spectroscopic analysis, our goal was to derive the monochromatic luminosities at different wavelengths and their emission lines properties, to test these quantities as $L_{UV}$ in the luminosities relation (Signorini+22A, in prep.). Up to now, the 2 keV and the 2500{\AA} monochromatic luminosities derived from photometric data have been used as $L_X$ and $L_{UV}$, respectively. These choices are rather arbitrary, as we are using two monochromatic luminosities as \textit{proxies} of the whole quasar emission. The second thing to notice is that, in the UV band, quasars show very strong emission lines. When we are using a photometric indicator, we are actually averaging the true quasar continuum, which is produced by the disc, and the lines contributions, which are produced in the BLR and in the NLR as a reprocessing of the central emission. The idea behind performing the spectral analysis was that if we manage to distinguish the continuum from the line contribution, we may get a lower dispersion of the luminosity relation.\\

\begin{figure*}
\centering
\includegraphics[scale=0.5]{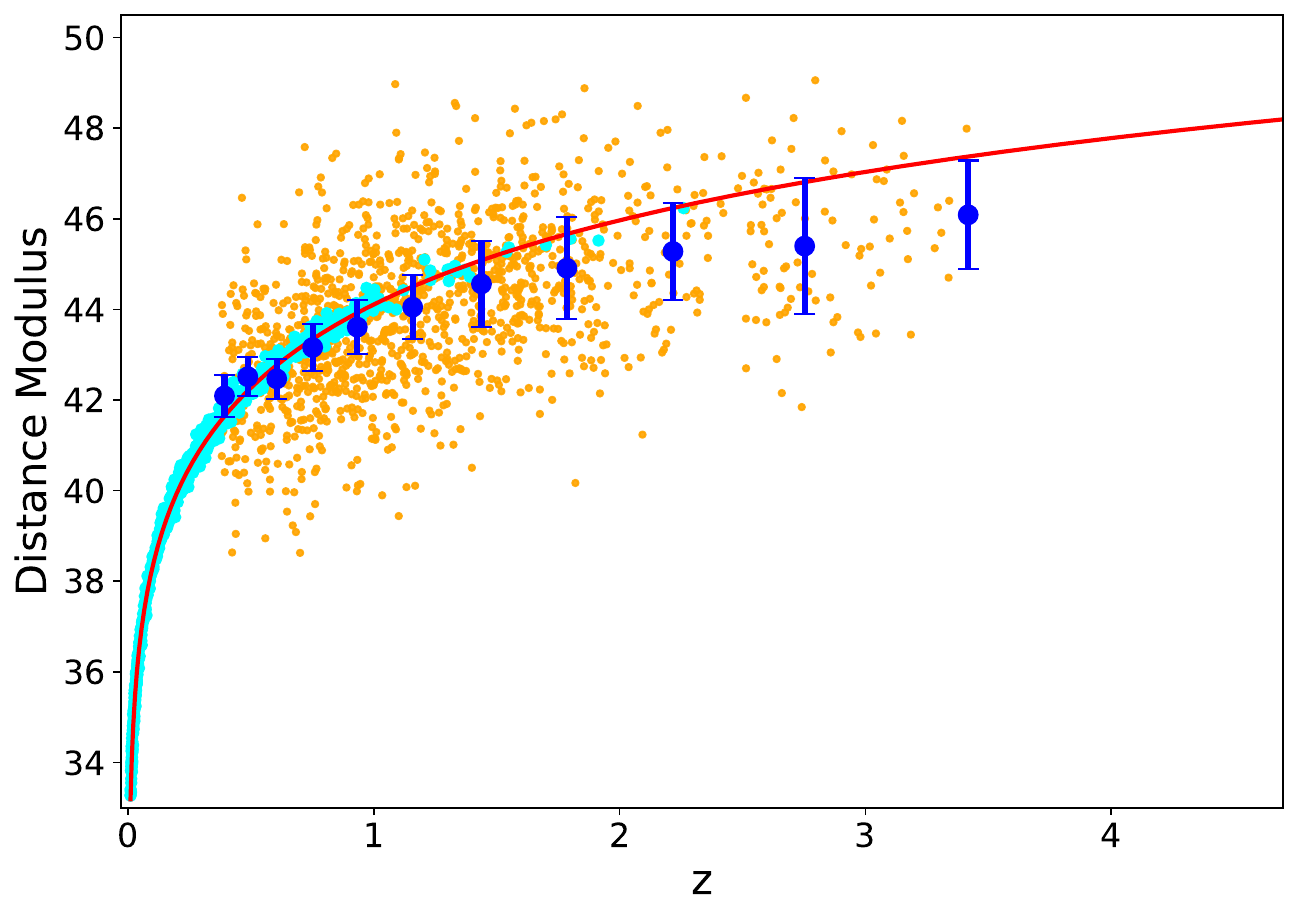}
\caption{Hubble Diagram built using the spectroscopically derived monochromatic 2500 {\AA} luminosities as $L_{UV}$. Yellow points show the Distance Modulus (DM) for quasars, blue points the average for quasars in small redshift bins, and cyan points the DM of SNIa. The red solid line is the flat-$\Lambda$CDM prediction. Although the relation itself doesn't have a smaller dispersion compared to when photometry is used, the different slope $\gamma$ means we get lower uncertainties on the distance estimates. This Hubble Diagram is consistent with the one obtained with photometry. The lower uncertainties make it so that the tension that is found with the predictions of the $\Lambda$CDM model is increased (Signorini+22A, in prep.).}
\label{fig:hd_spec}
\end{figure*}

When using the monochromatic luminosity at 2500 {\AA} derived from the UV spectroscopic analysis, which will be called $L_{spec}$, as $L_{UV}$, we obtain a relation with a slope $\gamma_{spec}=0.45\pm0.01$ and a dispersion $\delta_{spec}=0.22$ dex. This has to be compared with the results obtained using the photometric luminosity for the sample, $L_{phot}$, for which we get $\gamma_{phot}=0.60\pm0.01$ and $\delta_{phot}=0.22$ dex. \\
So, implementing luminosities derived from the spectroscopic analysis (i) does not give us a lower dispersion, and (ii) gives us a quite different average slope. We can interpret these results as follows: the spectroscopic and the photometric monochromatic luminosities are two different proxies of the quasar UV emission, as they take into account different properties of the emission itself. This is shown by the fact that we get different slopes when implementing one or the other. At the same time, they are two equally good proxies in terms of the dispersion we obtain by using them. \\
This result means we can not lower the uncertainties on the distance estimates due to the relation dispersion by using spectroscopic data. However, the fact that we get a lower slope $\gamma$, means that when we derive uncertainties on the luminosity distance $D_L$ from relation \ref{rel}, these will be smaller thanks to the $\frac{1}{2-2\gamma}$ factor being smaller. In this way, we can build the Hubble Diagram shown in Figure \ref{fig:hd_spec}. This Hubble Diagram is perfectly consistent with the one that is obtained with photometry for the same dataset (Figure 2 of \cite{RL19}). At the same time, the lower uncertainties make it so that the tension with the predictions of the $\Lambda$CDM model increases. Furthermore, this tension is found for a sample of objects whose UV properties have been accurately analysed, so we are sure the tension is not driven by the presence of reddening or other kinds of biases in the UV spectrum.\\

We also tested the luminosities relation using the luminosity of the MgII emission line as $L_{UV}$. The line is found at 2800 {\AA} and it is the line that is present for the largest number of objects in our sample, given the redshift range and the wavelength range of the observations. What we obtain is a slope $\gamma_{MgII}=0.60\pm0.01$, the same as the one we get with the photometry, and a lower dispersion, $\delta_{MgII}=0.18$ dex. The MgII emission, although the line is found at 2800 {\AA}, depends on the strength of the quasar emission at much shorter wavelengths, $\sim$800{\AA}, where the ionization is produced. This is a region of the quasar spectrum that is not directly observed due to absorption issues, but we can get a proxy of its intensity given the intensity of the line. The fact that we get a smaller dispersion might suggest that the physical relation is stronger at shorter wavelengths, where the peak of the quasar emission is also found. The slope difference between using the spectroscopic 2500{\AA} luminosity and the MgII luminosity can be explained as a consequence of the non-linear relation between the emission-line equivalent width (EW) and the luminosity of the quasar continuum, known as the Baldwin effect \citep{Baldwin77}. \\
As the objects in our sample all have UV data coming from the SDSS, we can consider that: 

\begin{equation}
\log({\rm EW_{Mg\,\textsc{ii}}}) = -0.214 \log(L_{3000}) + 11.388,
\label{BE}
\end{equation}

as derived by \cite{Rakshit20}. \\
By using equation \ref{BE} for the EW, we can then derive the Mg\,\textsc{ii} line luminosity as a function of the monochromatic continuum luminosity at 3000 {\AA} as:

\begin{equation}
\log(L_{\rm Mg\,\textsc{ii}}) = 0.786  \log(L_{3000}) + 11.388.
\label{mgg}
\end{equation}

If we now consider the relation between the X-ray and UV luminosities as $L_{2\,\rm keV} = \gamma L_{3000} + \beta$, with $\gamma = 0.45$, and we substitute $L_{3000}$ using equation \ref{mgg}, we obtain that $L_{2\,\rm keV} = \gamma^\prime L_{\rm Mg\,\textsc{ii}} + \beta^\prime$ with $\gamma^\prime = 0.45/0.786 = 0.59$, which is perfectly consistent with the slope of the relation that we obtain when using the Mg\,\textsc{ii} luminosity (or flux) as $L_{\rm UV}$. We conclude that the reason behind different slopes in the X-ray - UV relation when shifting from continuum UV proxies to line proxies is associated with the presence of the Baldwin effect itself. It is more difficult to give an explanation for the slope parameter that we obtain when using photometric UV fluxes, instead. The photometric fluxes are indeed a complex UV proxy, and they contain contributions from both the quasar continuum and line emissions.

\section{The Golden sample}
Another goal of the analysis of the $L_{UV}-L_{X}$ relation of the past few years has been the selection of a sample of objects with the highest possible quality both in the X-ray and in the UV, with the goal of verifying if with better quality data we do find a smaller dispersion. The results of these efforts have been recently published in the paper \cite{Sacchi22}. \\
The first step has been to perform a detailed spectral analysis on the X-ray side of the relation for all the objects in the \cite{Lusso20} sample with a redshift higher than 2.5, which are 130. The X-ray spectrum of a non-obscured quasar, which is the kind of object we are interested in, does not have any particular spectroscopic feature; we expect it to be a simple power-law. At the same time, the photon index of the power-law is a very important feature for this kind of analysis. We expect an unobscured quasar to have a photon index $\Gamma\sim1.8$, while objects with a lower value are probably obscured. This is one of the ``quality cuts" that we apply to the quasar sample to reduce the observed dispersion. Until now, we have been using photon indexes derived from photometric X-ray data. However, a complete X-ray spectral fit is surely better at obtaining a reliable estimate of the photon index and therefore at excluding from the sample objects with possible signs of obscuration. Unfortunately, the X-ray spectral reduction and analysis is very time-consuming, which is why we have only performed it for objects at redshift higher than 2.5.\\

\begin{figure}
\centering
\includegraphics[scale=0.22]{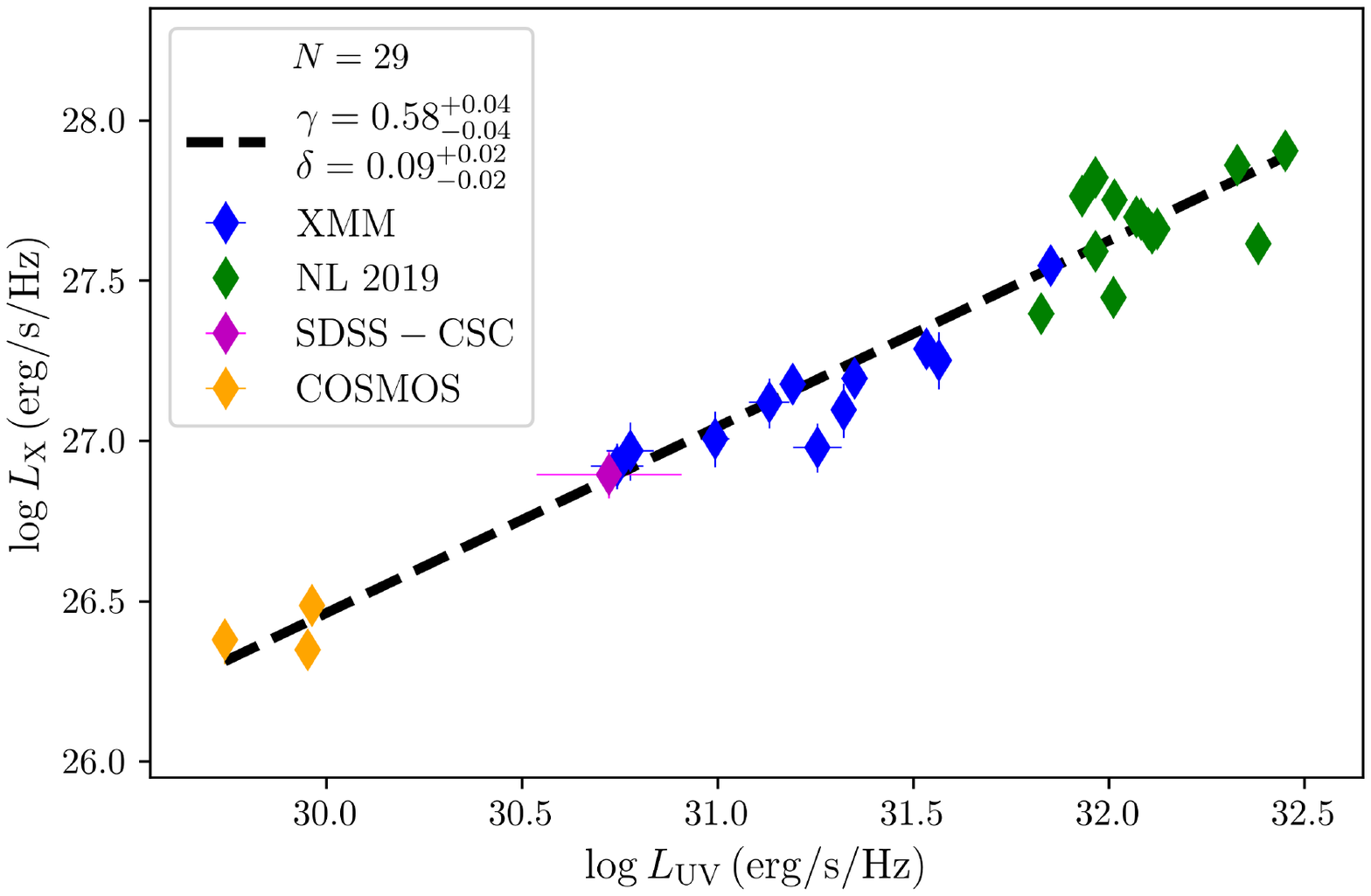}
\caption{$\log(L_X)-\log(L_{UV})$ for the $3.0 < z < 3.3$ “golden” sample. Colors refer to different observational subsamples. The fit of the X-ray - UV relation is done using fluxes instead of luminosities given that the objects are all found in a small redshift range. In this way, the results are also independent from cosmology. Figure from \cite{Sacchi22}.}
\label{fig:golden}
\end{figure}

Among these objects, which now have both UV and X-ray spectroscopic data analysed, we selected a ``Golden sample" of 30 objects at redshift $3<z<3.3$, among which a sample of 14 objects with direct, not serendipitous, X-ray observations are found \citep{Nardini19}. The objects in this sample do not have anything special, they simply have better-quality data. As can be seen in Figure \ref{fig:golden}, when we fit the $L_X-L_{UV}$ relation for these objects (which is done using fluxes instead of luminosities given that they are all found at similar redshifts), what we get is a very low dispersion, 0.09 dex \citep{Sacchi22}.

\begin{figure}
\centering
\includegraphics[scale=0.13]{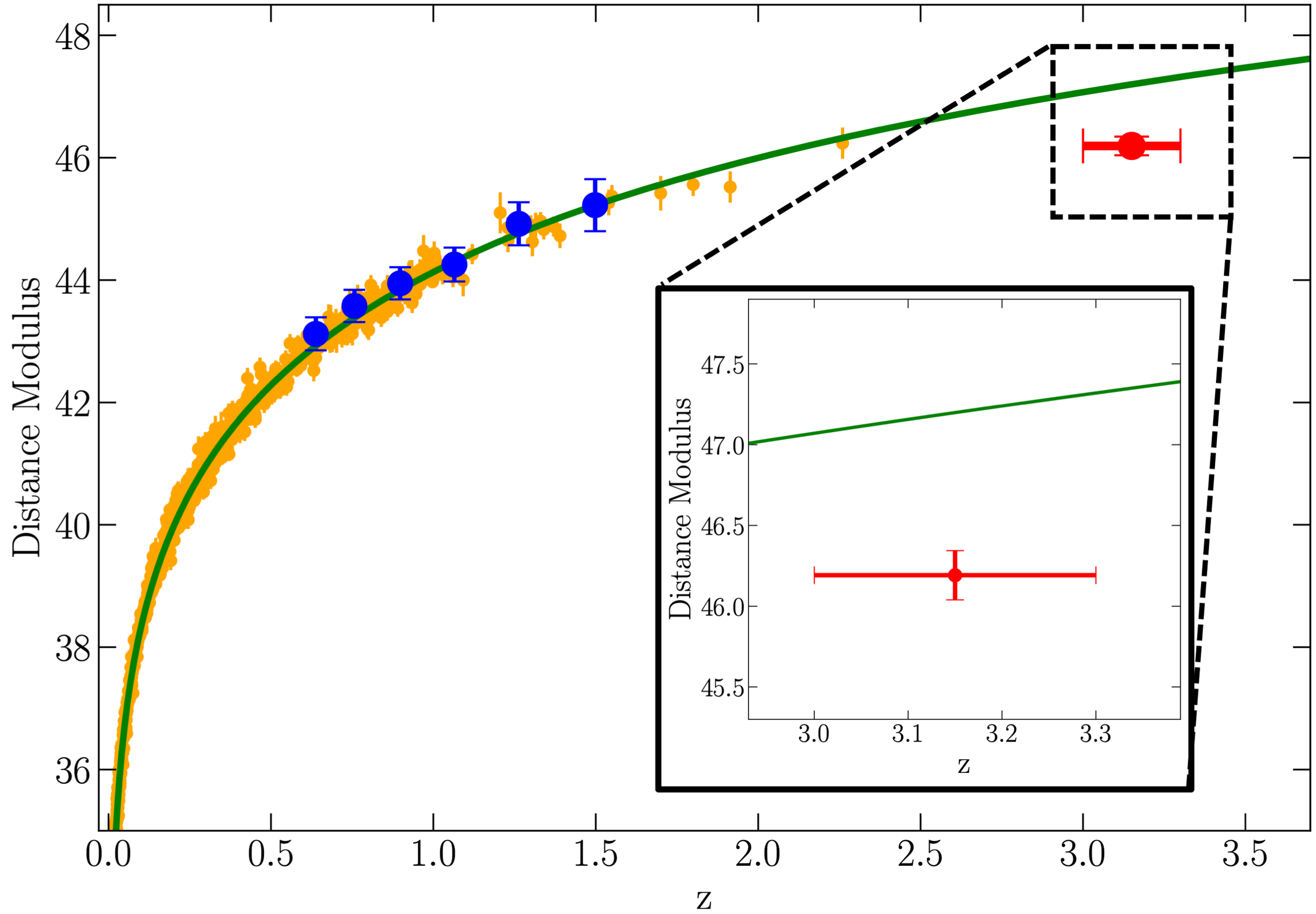}
\caption{Hubble Diagram of SNIa (orange) and quasars (blue), with the red point representing the "golden" sample at $3.0 < z < 3.3$. This DM  estimate is at a 5$\sigma$ tension with the flat $\Lambda$CDM prediction (green line). Figure from \cite{Sacchi22}.}
\label{fig:golden_HD}
\end{figure}

We know that this dispersion includes the contributions from some observational factors that can not be removed with our sample selection. These are the variability of quasars emission, the inclination with respect to the line of sight, and X-ray observational issues that might be reduced, but not eliminated, for objects with pointed X-ray observations. \\
From a recent analysis (Signorini+22B, in prep.) we found that these factors can fully explain the 0.09 dex dispersion. This means that the intrinsic dispersion of the physical $L_X-L_{UV}$ relation must be very low, possibly close to zero. The fact that intrinsically the dispersion must be so low gives reliability to our method and to the cosmological implementations of quasars as standard candles, although we are (still) not able to get such a low dispersion for all the objects in our sample. \\
For this subsample of objects, we can derive a distance measurement with a very small uncertainty, given the small dispersion of the relation. This distance is at a 5$\sigma$ distance from the prediction of the $\Lambda$CDM Hubble Diagram at redshift $z\sim3$, as can be seen from the red point in Figure \ref{fig:golden_HD}. This result confirms that the tension with the standard cosmological model is present and very significant, and it is found also for this subsample of objects which have the best quality data and whose UV and X-ray spectra have both been analyzed one-by-one.

\section{Conclusions}
The $L_X-L_{UV}$ relation has been shown to be a promising tool for cosmology, allowing us to extend the Hubble Diagram up to very high redshifts. The spectroscopic analysis of the UV side of the relation gives us a way to get more precise distance estimates, together with hints about the physics behind the relation itself. Restricting our analysis to a subsample of objects for which we have the best quality data shows an extremely low dispersion of 0.09 dex, which can completely be explained by observational factors. These results confirm the reliability of the implementation of quasars as standard candles, and the presence of a strong tension between the Hubble Diagram of quasars and the predictions of the flat-$\Lambda$CDM model, which might lead to new physics.


\end{document}